\documentclass[aps,prl,showpacs,twocolumn]{revtex4}
\usepackage{graphicx}
\usepackage{dcolumn}
\usepackage{bm}
\usepackage[latin1]{inputenc}
\usepackage{amsmath}
\usepackage{colordvi}
\usepackage{color}
\usepackage{ulem}

\begin{document}

\title{Relaxation of flying spin qubits in quantum wires by hyperfine interaction}

\author{C. Echeverr\'ia-Arrondo}
\affiliation{Departamento de Qu\'{\i}mica-F\'{\i}sica, UPV/EHU, Apdo 644, 48080 Bilbao, Spain}

\author{E. Ya. Sherman}
\affiliation{Departamento de Qu\'{\i}mica-F\'{\i}sica, UPV/EHU, Apdo 644, 48080 Bilbao, Spain}
\affiliation{Basque Foundation for Science IKERBASQUE, 48011 Bilbao, Spain}

\date{\today}

\begin{abstract}
We consider the relaxation of a spin qubit in a quantum dot propagating as a whole in a one-dimensional semiconductor with 
hyperfine coupling. We show that this motion leads to qualitatively new features
in this process compared to static quantum dots. For a fast straightforward motion, the initial spin density
decreases to zero with the relaxation rate independent of the spatial spread 
of the electron wave function and inversely 
proportional to the electron speed. 
However, for the oscillatory motion, the qubit acquires memory, and the dephasing becomes
Gaussian rather than exponential. After some time, one third of 
the initial spin polarization is restored, as it happens for static qubits. 
This revival can occur either through periodic peaks  
or through a monotonous increase in the polarization, after a minimum, until a plateau
has been reached. Our results can be useful for the understanding of the 
spin dynamics and decoherence in quantum wires.

\end{abstract}

\pacs{72.25.Rb,73.21.Hb,31.30.Gs,03.67.-a}

\maketitle


Semiconductor quantum wires play an important role in fundamental and 
applied spintronics. They allow one to study a variety of interesting phenomena related to spin transport 
\cite{Pramanik,Kiselev,Quay,Bringer,Governale,Pershin,Holleitner,Sanchez,Lu,Romano,Japaridze}  
and open the way to realizations of exotic states such as  
Majorana fermions \cite{Mourik12,Gangadharaiah}. 

One of the most interesting aspects of spintronics is the hyperfine coupling of spins 
of carriers to the nuclear magnetic moments of the host lattice
\cite{Glazov12,Greilich,Coish,LianAo,Marcus1,Dahbashi}, strongly dependent
on the spatial extension of the carriers wave functions. 
In semiconductor nanostructures, it is enhanced compared to the bulk due 
to confined wave functions in quantum dots \cite{Merkulov,Braun,Raith} 
and due to localization by disorder in quantum wires.\cite{Echeverria2} The spin
dephasing times observed in quantum dots \cite{Braun} agree very well with 
what can be expected from the theory.\cite{Merkulov}

The observed relaxation times suggest that a single electron spin could become the physical 
realization of a quantum bit (qubit), crucial for quantum information 
applications. Unlike the dots, where electrons are well localized, 
extended nanowires are natural channels for traveling qubits, hereafter 
referred to as "flying" qubits \cite{Yamamoto,Nadj}. Spin decoherence is the major concern for
using qubits in information processing. At low temperatures, it is mainly 
caused by spin-orbit and hyperfine interactions. In undoped 
wires, the mechanisms for spin-orbit relaxation can be canceled out by 
tuning their cross section geometries for the wires extended along the 
[001] direction (Dresselhaus term) and also by keeping their environments 
symmetric (Rashba term). Hyperfine couplings 
hence can become the only source of spin depolarization. Their influence on the spin 
dynamics of qubits propagating through nanowires should be deeply understood for 
spin transport devices. Recently, Huang and Hu \cite{Hu} have proposed a 
formulation of the spin-relaxation problem well suited for quantum information 
purposes. This formulation comes from studying the decoherence of an electron spin 
that is being transported in a random electric potential, as enclosed in a quantum dot with the 
time-independent shape of the wave function. The ability of high-fidelity electron transfer
between quantum dots by surface acoustic waves  was demonstrated recently.\cite{Stotz,Hermelin,McNeil,Sanada} 
Here we apply a similar approach to study the effects of hyperfine interaction
on spins in moving quantum dots, and show that the dynamics brings about new 
time scales and qualitative features compared to the static dots. 

\begin{figure}[tbp]
\begin{centering}
\includegraphics*[scale=.45]{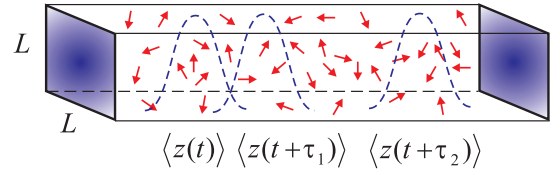}
\caption{(Color online) Schematic plot of a semiconductor quantum wire. 
Random arrows show the magnetic moments of nuclei.
Dashed lines correspond to density distributions with given
expectation values of the $z$ coordinate at different times:
For $\tau_{1}$, of the order of $\tau_{\xi}$, the matching of tails yields some correlation; 
for $\tau_{2}$, considerably larger than $\tau_{\xi}$, correlation is negligible.}
\label{fig:fig1}
\end{centering}
\end{figure}

We consider nanowires with $L\times L$ cross section, 
see Fig.~\ref{fig:fig1}, and we take GaAs as a representative spintronics material. 
Nuclear magnetic moments are considered frozen on the time scale of the 
electron precession \cite{Merkulov}, and are represented as random $I=3/2$ spins
with the expectation values of components $I_i$, where $i=x,y,z$. 
Under the assumption of decoupled spin and orbital motion, valid
in the absence of spin-orbit coupling and for weak hyperfine interaction,
we can factorize the total wave function as $\psi (\mathbf{r},t)\left|\phi\rangle\right.$.
Here the orbital state is $\psi (\mathbf{r},t)$ and the spinor state is $\left|\phi\rangle\right.$. 
The resulting Hamiltonian for the spin degree of freedom depends on the orbital wave function 
as follows:
\begin{equation}
\hat{H}_{\text{hf}}=
\frac{A\nu}{2}\sum_{n}|\psi (\mathbf{r}_n,t)|^{2}(I_{x}^{[n]}\sigma _{x}+I_{y}^{[n]}\sigma_{y}+I_{z}^{[n]}\sigma_{z}),
\label{Hhf}
\end{equation}
where the summation goes over all nuclei with spin components
$I_{i}^{[n]}$  at sites $\mathbf{r}_{n}$, $A$= 45 $\mu $eV is the typical 
hyperfine coupling constant for GaAs \cite{Merkulov}, 
$\nu $ is the volume per single nucleus (one eighth of the unit cell), and $\sigma_{i}$ are the Pauli matrices.

Hamiltonian (\ref{Hhf}) can be rewritten in the form 
\begin{equation}
\hat{H}_{\text{hf}}=\frac{\hbar }{2}\left({\bm\Omega}\cdot{\bm\sigma}\right),
\end{equation}
where the precession due to hyperfine interaction $\mathbf{\Omega}$ is an integral that depends on 
the electron position and can be expressed as
\begin{equation}
\Omega _{i}=\frac{A\nu}{\hbar} \int_V |\psi (\mathbf{r},t)|^{2}\rho _{i}(\mathbf{r}) dV.
\end{equation}
Here the density of the $i$-th component of
the nuclear magnetic moment $\rho _{i}(\mathbf{r},t)$  satisfies the white-noise
distribution $\left\langle\langle \rho _{i}(\mathbf{r})\rho _{i}(\mathbf{r}^{\prime})\right\rangle\rangle
=\rho _{0}\delta \left( \mathbf{r-r}^{\prime }\right),$
where $\langle\langle \ldots \rangle\rangle$ stands for the ensemble 
average and $\rho_{0}=\langle\hat I_{i}^2 \rangle/\nu$; here $\nu^{-1}$ 
is the concentration of nuclei. To describe a moving electron, we need the characteristic time-dependent correlator 
which can be obtained, assuming Gaussian random fluctuations \cite{Shklovskii,Glazov05}, as: 
\begin{eqnarray}
&&\frac{\hbar^2}{\left(A\nu\right)^{2}}\left\langle\langle \Omega _{i}(t)\Omega _{i}(t+\tau )\right\rangle\rangle =
\label{OmOm}\\ 
&&\int\int |\psi (\mathbf{r}_{1},t)|^{2}|
\psi(\mathbf{r}_{2},t+\tau )|^{2}\left\langle\langle \rho _{i}(\mathbf{r}_{1})\rho_{i}
(\mathbf{r}_{2})\right\rangle\rangle dV_{1}dV_{2}\nonumber\\
&&=\rho _{0}\int\int|\psi (\mathbf{r},t)|^{2}|\psi(\mathbf{r},t+\tau)|^{2}dV. \nonumber
\end{eqnarray}
Although, due to the isotropy of the random distribution of nuclear spins, 
these correlators are $i-$independent, for definiteness we consider the $z-$ component 
of the spin below. With the course of time, the overlap of the electron wave function 
with itself decreases and the correlation gradually vanishes, 
as can be realized from Fig.\ref{fig:fig1}. The $\psi (\mathbf{r}_{n},t)$ states characterizing 
the propagating qubits are defined as 
$\psi (\mathbf{r}_{n},t)=\eta (z_{n},t)\varphi (x_{n},y_{n})$, where $\varphi
(x_{n},y_{n})=(2/L)\sin (\pi x_{n}/L)\sin (\pi y_{n}/L)$ describes an electron in a single-mode
box, and $\eta (z_{n},t)$ is assumed to be a Gaussian centered at the time-dependent position $\langle z(t)\rangle$, 
corresponding to the ground state of the moving parabolic potential:
\begin{equation}
\eta(z_n)=\left(\frac{1}{2\pi\xi^2}\right)^{1/4}\exp\left[{-(z_n-\langle z(t)\rangle)^2/4\xi^2}\right].
\end{equation}
For this choice of the wave function, the correlator in Eq.(\ref{OmOm}) is expressed as 
\begin{equation}
\left\langle\langle \Omega _{i}(t)\Omega _{i}(t+\tau )\right\rangle\rangle =
\left\langle\langle\Omega _{i}^{2}\right\rangle\rangle \exp \left[ -\left(\Delta z(\tau)\right)^{2}/4\xi^{2}\right], 
\end{equation}
where $\langle\langle\Omega _{i}^{2}\rangle\rangle=3\left(A/\hbar\right)^{2}I(I+1)\nu/8\sqrt{\pi}L^{2}\xi$ 
and $\Delta z(\tau)\equiv\langle z(t+\tau) \rangle - \langle z(t) \rangle$. 
This correlator reaches the maximum when the electron is back to the position occupied at time $t$. 
For motion with constant speed, $\Delta z(\tau )=v\tau$ and the corresponding
effective correlation time can be defined as $\tau_{\xi}\equiv \xi /v$.

Now we analyze the evolution of polarization of static and propagating wave
packets.  The spin dynamics is calculated for ensembles of random wires and
averaged afterwards. At the initial time, qubits are fully polarized in the
nonequilibrium spin state $|\phi (0)\rangle =|1\rangle $. Later on, for 
$t>0$, $|\phi (t)\rangle $ evolves following the time dependent Schr\"{o}dinger
equation,
\begin{equation}
i\hbar \frac{\partial \left|\phi\right> }{\partial t}=\hat{H}_{\text{hf}}(t)\left|\phi\right> ,
\end{equation}
where the time dependence of $\hat{H}_{\text{hf}}(t)$ is due to the displacement of the qubit. 
The resulting spinor $|\phi (t)\rangle$ is related to the initial spin state as
\begin{equation}
|\phi (t)\rangle =\mathcal{T}\exp \left[ -\frac{i}{\hbar }\int_{0}^{t}\hat{H}%
_{\text{hf}}(t)dt\right] |\phi (0)\rangle,
\end{equation}
where $\mathcal{T}$ stands for the time ordering.

\begin{figure}[tbp]
\begin{centering}
\includegraphics*[scale=.4]{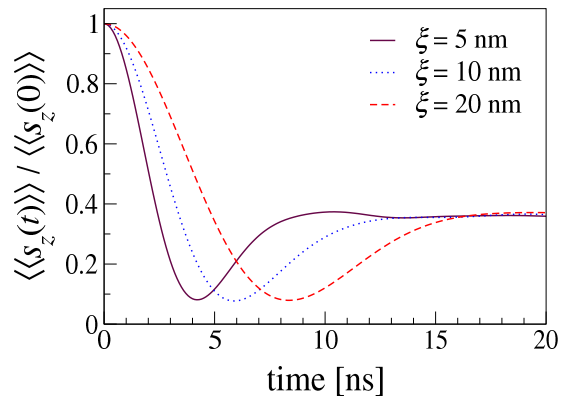}
\caption{(Color online) Spin dynamics of static qubits for different
width $\xi$ as solved by $4^{\mathrm{th}}-$order Runge-Kutta method
and averaged over 4096 realizations of disorder in the nuclear magnetic moments
with the statistical errors of the order of $1/\sqrt{4096}=1.6\%$.  
Here and below we assume the wire width $L=8$ unit cells.  For this wire 
width and $\xi=10$ nm, we obtain $1/\sqrt{\langle\langle\Omega_{i}^{2}\rangle\rangle}\approx 1.5$ ns,
corresponding well to the relaxation time in the Figure.
Since the GaAs lattice constant is $0.565$ nm, the nanowire cross-section is 
$4.5\times 4.5\simeq 20$ nm$^2$.}
\label{fig:fig2}
\end{centering}
\end{figure}


We consider three different realizations of spin dephasing: for static qubits (to have the known
system to compare with), and for straightforward and oscillatory
motion. When packets are fixed to initial positions, the calculated average polarization decays to a minimum of
about 0.1 of the initial value and monotonically increases up to a constant
plateau where $\langle\langle s_{z}(t)\rangle\rangle =\langle\langle s_{z}(0)\rangle\rangle/3$.
We consider packets with widths of $\xi =5$, 10, and 20 nm; 
their spin evolutions, plotted in Fig.~\ref{fig:fig2}, look exactly as expected from the 
dynamics induced by hyperfine coupling in quantum dots \cite{Merkulov, Echeverria2}. 
This steady polarization arises due to the 
precessional motion of the qubit spin around the time-independent 
effective field of the nuclear magnetic moments \cite{Merkulov}. 
The spin evolution presented in this Figure is universal in the
sense that it can be fully described by two timescales originated from the
same random position-dependent spin precession:
that for the initial polarization to decay down to $1/e$, and that 
for reaching the 1/3 plateau. 
The scale for the initial decay rate of static electrons 
is then $\Gamma_{\rm st}=\sqrt{\langle\langle\Omega_{i}^{2}\rangle\rangle}$, 
and the entire time dependence in Fig.~\ref{fig:fig2} can be described by 
using this single parameter \cite{Merkulov}. The dependence of the spin dynamics on the 
qubit spatial width is related to the 
number of interacting nuclei inside the electron cloud: A smaller amount of nuclei yields 
stronger fluctuations in the field resulting in a faster decay.

\begin{figure}[tbp]
\begin{centering}
\includegraphics*[scale=.4]{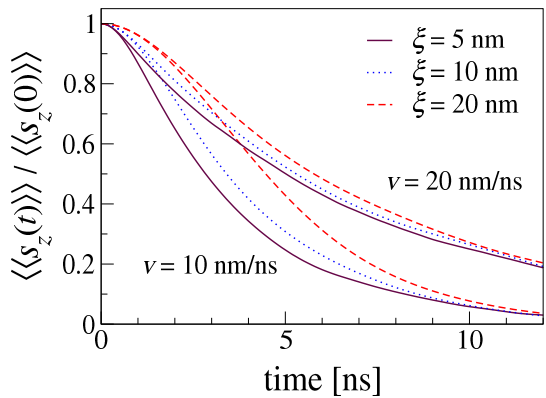}
\caption{(Color online) Average polarization for qubits in straightforward 
motion for different speeds $v$ and widths $\xi$.}
\label{fig:fig3}
\end{centering}
\end{figure}

In the next, straightforward motion regime, packets propagate along the wire with
constant velocities $v=10$ and $20$ nm/ns, taken here as examples. A behavior considerably different
from the static one emerges when $\tau_{\xi}\Gamma_{\rm st}$ is much less than 1, that
is, when the electron moves to a strongly different realization of the hyperfine field before 
sufficiently changing its spin; this behavior is shown in Fig.~\ref{fig:fig3}. 
To gain insight into the
influence of width and speed, we have taken into account, again, three
distinct packet sizes, $\xi =5$, 10, and 20 nm. When $v=10$ nm/ns all polarizations decrease below 0.1 after
10 ns and monotonically tend to zero afterwards. In addition, the smaller the
qubit width, the faster the dephasing. Similar calculations for $v=20$
nm/ns show a considerably less efficient decoherence. This behavior can be detailed as follows. 
The average spin evolution can be described with \cite{Glazov05} 
\begin{equation}
\frac{\left\langle\langle s_{z}(t)\right\rangle\rangle}{\left\langle\langle s_{z}(0)\right\rangle\rangle}=
\exp
\left[ -\int_{0}^{t}dt^{\prime }\int_{0}^{t^{\prime }}
\langle\langle 
\Omega_{z}(t^{\prime })\Omega_{z}(t^{\prime }-t^{\prime\prime })
\rangle\rangle dt^{\prime \prime}\right].
\end{equation}
For the exponentially decaying part of the dynamics, this expression has the form
\begin{equation}
\frac{\left\langle\langle s_{z}(t)\right\rangle\rangle}{\left\langle\langle s_{z}(0)\right\rangle\rangle}
= \exp (-\sqrt{\pi}\Gamma_{v}t), 
\end{equation}
where the spin relaxation rate is determined by
\begin{equation}
\label{gamma}
\Gamma_{v} =\langle\langle{\Omega}_{i}^{2}\rangle\rangle\tau_{\xi}. 
\end{equation}
Therefore, the fast motion establishes a new and smaller relaxation 
rate, $\Gamma_{v}=\Gamma_{\rm st}\cdot\left(\Gamma_{\rm st}\tau_{\xi}\right)$, 
such as $\Gamma_{v}\ll \Gamma_{\rm st}$.
Here $\Gamma_{v}$ does not depend on the width of the electron wave packet but only on its speed, 
as we explain below using a scaling argument. The electron spin precession rate, due to
the frozen nuclear spins, is of the order of $\Omega \sim \left(A/\hbar\right)/\sqrt{N},$ 
where the number of nuclear magnetic moments per single electron is $N\sim L^{2}\xi /\nu$.
From Eq.(\ref{gamma}), the resulting $\Gamma_{v}$ is proportional to $\left(A/\hbar \right)^{2}\nu /L^{2}v$ and, henceforth, 
also $\xi -$independent. This argument can easily be applied to other spatial dimensions $D$ in the following way. 
The square of the spin precession rate being of the order of 
$\langle\langle\Omega_{i}^{2}\rangle\rangle \sim
N^{-1}\sim \xi^{-D},$ while $\tau \sim \xi /v$, is $D-$independent. As a
consequence, $\Gamma_{v}\sim\xi ^{1-D}/v$, which tends to zero for bulk crystals ($D=3$) and
two-dimensional structures ($D=2$) when the size of the wave packet is 
increased. However, it remains constant in the wires ($D=1$), making the results only weakly dependent
on the details of the electron wave function, provided that it broadens relatively slow, on the 
timescale of spin precession in the hyperfine field, such that the fast motion condition 
$\tau_{\xi}\Gamma_{\rm st}\ll 1$ holds. We notice that, for the electrons transferred 
by the surface acoustic waves \cite{Stotz,Hermelin,McNeil,Sanada} with the speed close to $3\times10^{3}$ nm/ns, 
the condition of fast motion is well satisfied for $\xi$ up to 500 nm. As a result, the spin states are 
well protected against relaxation.

\begin{figure}[tbp]
\begin{centering}
\includegraphics*[scale=.4]{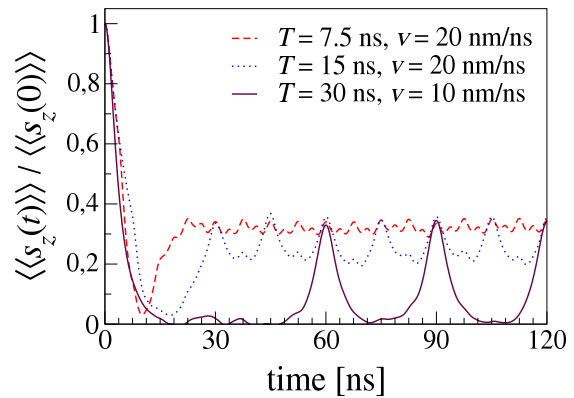}
\caption{(Color online) Average polarization for qubits flying back and forth along
the wires for various speeds and periods, $\xi=10$ nm. Note the sharp revivals in the regime 
when $T=30$ ns and $v=10$ nm/ns.}
\label{fig:fig4}
\end{centering}
\end{figure}

A significantly different behavior is obtained when the packet moves back and
forth,  in a regime which can be experimentally achieved  by the surface acoustic wave
technique.\cite{McNeil} We analyze oscillations with different velocities
and frequencies, which make amplitudes in the range of several tens of nanometers.
We assume that the expectation value of the qubit $z-$coordinate follows a saw-like function,
\begin{equation}
\left\langle z(t)\right\rangle =
2\ell\left| \frac{t}{T}-\left\lfloor \frac{t}{T}+\frac{1}{2}\right\rfloor \right|,
\end{equation}
where $\lfloor\ldots\rfloor$ stands for the floor function, $T$ is the period, and $\ell=vT/2$ 
is the maximum displacement from the initial position. 
Every time the electron returns to the initial position, 
once per period, it scans exactly the same effective field. 
As a consequence, the spin dynamics experiences strong correlations 
which yield the memory effects. In the oscillatory regime, 
the correlator for the precession rate satisfies the conditions $\left\langle\langle
\Omega _{i}(t)\Omega _{i}(T(\left\lfloor 2t/T\right\rfloor
+1)-t)\right\rangle\rangle =\left\langle\langle \Omega _{i}^{2}\right\rangle\rangle$ and $%
\left\langle\langle \Omega _{i}(t)\Omega _{i}(t+nT)\right\rangle\rangle =\left\langle\langle
\Omega _{i}^{2}\right\rangle\rangle$, where $n\ge 0$ is an integer. 

We begin with the regime where $T\Gamma_{v}$ is of the order of 1 and
the typical electron spin strongly changes in a single oscillation. 
The spin polarization behavior is shown in Fig.~\ref{fig:fig4}. 
We begin with the period $T=7.5$ ns and velocity $v=20$ nm/ns, where 
spin relaxation during half a period is not very strong yet. As a result, the
polarization first decays and then increases to a stable
value with tiny saw-like peaks on the top.
For a twice larger period, $T=15$ ns and $v=20$ nm/ns, the 
product $T\Gamma_{v}$ is larger by a factor of 2, spin relaxation
per period is considerably more efficient, and the
calculated dynamics already presents clear periodic peaks above the reached plateau.
For a yet smaller frequency and speed, $T=30$ ns and $v=10$ nm/ns, 
one observes an almost flat stage followed by a series of sharp
revivals periodically repeated in time, see Fig.~\ref{fig:fig4}. Here, in the first period,
the polarization relaxes down to zero. Throughout the second period, the qubits show 
a revival to 1/3 of the initial spin. Beyond
two periods, relaxation and memory effects both contribute to yield a
periodic picture of revivals up to one third and decays down to zero. 

\begin{figure}[tbp]
\begin{centering}
\includegraphics*[scale=.4]{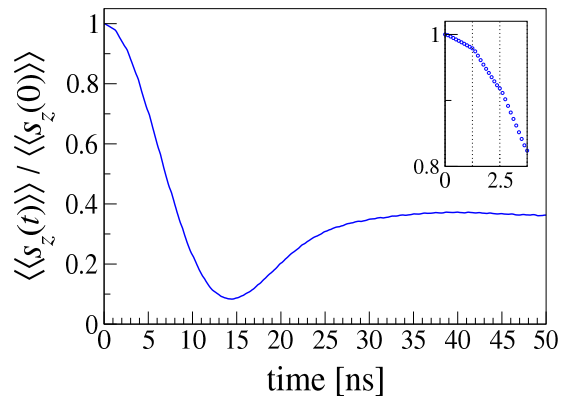}
\caption{(Color online) Average polarization of an oscillating qubit  
when $T=2.5$ ns, $v=160$ nm/ns, and $\xi=10$ nm. The inset shows details of
the short-term behavior changing slope every half a period shown by vertical dotted lines.}
\label{fig:fig5}
\end{centering}
\end{figure}

In the next case, we consider very high velocity and frequency, $T\Gamma_{v}\ll1$, resulting in
a small spin precession angle in one period. When $v=160$ nm/ns and $T=2.5$ ns, the average spin polarization 
decays to a minimum of about 10\% of
the initial value and monotonically increases up to a constant plateau of
1/3; the ensemble dephasing time is roughly 14 ns, as can be seen in
Fig.~\ref{fig:fig5}. Due to the extremely high speed, the qubit spin
cannot follow, by precessional motion, the fast changes in the
effective nuclear magnetic field. The characteristic behavior is given in Fig.~\ref
{fig:fig5}, similar to that depicted in Fig.~\ref{fig:fig2} for static qubits. 
Here, the behavior of polarization in time cannot be modified by tuning the qubit width.

In the oscillatory regime, the relaxation part of the time-dependence is 
Gaussian rather than exponential \cite{Glazov05,Reuther}, as we explain below. 
At any $t$, ensemble mean value of spin $z-$ component is $\left\langle\langle \cos \theta _{z}\right\rangle\rangle/2$, where
$\theta_{z}$ is the corresponding precession angle; 
this mean value is $\left\langle\langle \cos \theta _{z}\right\rangle\rangle =\exp(-\sqrt{\pi}\Gamma_{v}t)$ for $0<t<T/2$, and is multiplied by factor 
$n$, $\left\langle\langle \cos \theta _{z}\right\rangle\rangle =\exp
(-\sqrt{\pi}n^{2}\Gamma_{v}t)$, at $t=nT/2$. Thus, in this regime, a new spin relaxation 
rate  is established, $\Gamma_{\rm mem}=\sqrt{\Gamma_{v}/T}\gg\Gamma_{v}$, similar
to the memory effects in the systems with random spin-orbit coupling subject to
external magnetic field \cite{Glazov05,Glazov10}.
On the other hand, since $\Gamma_{\rm mem}=\Gamma_{\rm st}\sqrt{\tau_{\xi}/T}\ll\Gamma_{\rm st}$, this
relaxation is found even slower than for static qubits. Moreover, due to the memory effects, 
the slopes of $\ln\left(\langle\langle s_{z}(t)\rangle\rangle/\langle\langle s_{z}(0)\rangle\rangle\right)$
at $t=nT/2$ should change following the ratios $(2n+1)/(2n-1)$, which are indeed very 
close to those obtained from our numerical data, see inset of Fig.~\ref{fig:fig5}. 

To summarize, we have investigated the spin dynamics due to hyperfine coupling
of electrons embedded in quantum dots propagating along quantum wires. 
This behavior of spins is qualitatively different from the typical single-parameter time dependence
in static dots. For straightforward motion with constant velocity,
the spin relaxation is close to exponential towards zero and,
due to motional narrowing, its rate does not depend
on the spatial width of the packets but solely on their speed.
On the contrary, in the oscillatory regime, the decay
can be Gaussian rather than exponential,
and the polarization revives afterwards up to one third of the initial value.
Two modes characterize the oscillatory regime: 
The spin revival can be reached through periodic peaks and valleys, and,  
in the other mode, through a monotonous increase of 
polarization towards a stable plateau. The shown dynamic trends, especially interesting 
for oscillating qubits, adds to the knowledge of the behavior of "flying" electrons in 
semiconductor nanowires. They also broaden the applicability of such nanosystems in the 
fields of spintronics and quantum information technology.

This work was supported by the MINECO of Spain grant FIS2012-36673-C03-01,
"Grupos Consolidados UPV/EHU del Gobierno Vasco" grant IT-472-10, and by the
UPV/EHU under program UFI 11/55. We are grateful to Xuedong Hu for valuable discussions.

\end{document}